\documentclass[%
reprint,
superscriptaddress,
 amsmath,amssymb,
 aps,
 prl,
]{revtex4-2}

\usepackage{graphicx}
\usepackage{dcolumn}
\usepackage{bm}
\usepackage{hyperref}
\usepackage[mathlines]{lineno}
\usepackage{comment}

\usepackage{dcolumn}
\usepackage{multirow}
\usepackage{booktabs}
\usepackage{xcolor}
\usepackage{ulem}

\usepackage{lineno}
\usepackage{amsmath}
\usepackage{etoolbox} 

\newcommand*\linenomathpatch[1]{%
  \cspreto{#1}{\linenomath}%
  \cspreto{#1*}{\linenomath}%
  \csappto{end#1}{\endlinenomath}%
  \csappto{end#1*}{\endlinenomath}%
}
\linenomathpatch{equation}
\linenomathpatch{gather}
\linenomathpatch{multline}
\linenomathpatch{align}
\linenomathpatch{alignat}
\linenomathpatch{flalign}

\newcommand{\APV}{A_{\rm PV}}

\newcommand{\AT}{A_{\rm T}}
\newcommand{\pb}{^{208}{\rm Pb}}
\newcommand{\eca}{^{48}{\rm Ca}}

\newcommand{\FW}{F_{\rm W}}
\newcommand{\Fch}{F_{\rm ch}}
\newcommand{\RW}{R_{\rm W}}
\newcommand{\Rch}{R_{\rm ch}}

\begin{document}

\preprint{APS/123-QED}

\title{Precision Determination of the Neutral Weak Form Factor of $^{48}$Ca}

\collaboration{The CREX Collaboration}

\author{D.~Adhikari}\affiliation{Idaho State University, Pocatello, Idaho 83209, USA}
\author{H.~Albataineh}\affiliation{Texas  A  \&  M  University - Kingsville,  Kingsville,  Texas  78363,  USA}
\author{D.~Androic}\affiliation{University of Zagreb, Faculty of Science, Zagreb, HR 10002, Croatia}
\author{K.A.~Aniol}\affiliation{California State University, Los Angeles, Los Angeles, California  90032, USA}
\author{D.S.~Armstrong}\affiliation{William \& Mary, Williamsburg, Virginia 23185, USA}
\author{T.~Averett}\affiliation{William \& Mary, Williamsburg, Virginia 23185, USA}
\author{\mbox{C. Ayerbe Gayoso}}\affiliation{William \& Mary, Williamsburg, Virginia 23185, USA}
\author{S.K.~Barcus}\affiliation{Thomas Jefferson National Accelerator Facility, Newport News, Virginia 23606, USA} 
\author{V.~Bellini}\affiliation{Istituto  Nazionale  di  Fisica  Nucleare,  Sezione  di  Catania,  95123  Catania,  Italy}
\author{R.S.~Beminiwattha}\affiliation{Louisiana Tech University, Ruston, Louisiana 71272, USA}
% ...

\date{\today}

\begin{abstract}
We report a precise measurement of the parity-violating asymmetry 
$\APV$ in the elastic scattering of longitudinally polarized electrons from $\eca$. 
We measure 
$\APV =2668\pm 106\ {\rm (stat)}\pm 40\ {\rm (syst)}$ parts per billion, 
leading to an extraction of the neutral weak form factor $\FW (q=0.8733$ fm$^{-1})
= 0.1304 \pm 0.0052 \ {\rm (stat)}\pm 0.0020\ {\rm (syst)}$ and the charge minus the weak form factor $\Fch - \FW = 0.0277\pm 0.0055$. The resulting 
neutron skin thickness $R_n-R_p=0.121
\pm 0.026\ {\rm (exp)} \pm 0.024\ {\rm (model)}$~fm is relatively
thin yet consistent with many model calculations. 
The combined CREX and PREX results will have implications for future energy density functional
calculations and on the density dependence
of the symmetry energy of nuclear matter. 

\end{abstract}

\maketitle
Parity-violating electron scattering (PVES) can  locate neutrons in nuclei with minimal model dependence since the electroweak reaction is free from most strong interaction uncertainties \cite{Donnelly:1989,Horowitz:2001,Thiel:2019tkm}. PVES measurements can be optimized to extract the thickness of the neutron skin, the excess in the root mean-square size of the distribution of neutrons over that of the protons,  which depends on the pressure of neutron-rich matter as neutrons are pushed out against surface tension \cite{Brown:2000}. Recently PREX-2 accurately measured the thickness of the neutron skin in $^{208}$Pb using this technique~\cite{PREX:2021umo}.

Chiral effective field theory can predict neutron skin thicknesses using two- and three-nucleon interactions~\cite{Hagen:2016}. These interactions are typically measured in few-nucleon systems where important  three-neutron forces~\cite{Bentz:2020mdk} are difficult to probe.
Although such calculations using coupled cluster wave functions for both $\eca$ and $^{208}$Pb have now been performed~\cite{Hu:2021,Hagen:2016},
microscopic calculations are more feasible in the lighter $\eca$ system than for $^{208}$Pb.  We report here on a PVES measurement to constrain the neutron radius of $\eca$. 
While the $^{208}$Pb nucleus more closely approximates uniform nuclear matter, the $\eca$ nucleus lies in a different regime of smaller nuclei for which the neutron skin is more closely related to the details of the nuclear force. Not only is the new measurement complementary to the earlier $^{208}$Pb result in this way, but it will allow direct comparison to more microscopic calculations.

More accurate neutron skin predictions across the periodic table~\cite{Novario:2021low,PhysRevC.88.031305,arxiv.2203.09753} will be facilitated by these measurements in $\eca$ and $\pb$.
Since atomic parity violation experiments depend on the overlap of atomic electrons with neutrons, PVES neutron radii constraints along with nuclear theory may allow more precise low energy tests of the Standard Model \cite{APV_Cs, Bouchiat_Rev, Ramsey_Musolf_Rev1, APV_Safronova}. Coherent neutrino-nucleus elastic scattering depends on neutron radii and the same weak form factor as does PVES~\cite{NuCoherent1, NuCoherent2}. PVES weak form factor measurements along with theory may improve sensitivity to non-standard neutrino interactions.
A neutron star is 18 orders of magnitude larger than a heavy nucleus yet they have similar density, and both systems are governed by the same strong interactions and equation of state relating pressure to density.~\cite{Thiel:2019tkm,Novario:2020kuf,Shen:2020sec,Horowitz:2019piw,Wei:2019mdj}. Therefore, laboratory neutron skin measurements have important implications for neutron star properties, such as radius and tidal deformability \cite{Fattoyev:2018}, and are complementary to direct X-ray~\cite{Riley:2019yda} and gravitational wave observations~\cite{Chatziioannou:2020pqz,Zhang:2020azr,Guven:2020dok,Baiotti:2019sew,Piekarewicz:2019ahf,Tsang:2019mlz,Fasano:2019zwm}.  

Information on the $\eca$ weak charge distribution is obtained by measuring the PVES asymmetry ($\APV$) of longitudinally polarized 
electrons off an isotopically enriched  $\eca$ target 
in Hall A at Thomas Jefferson National Accelerator Facility (JLab). At first Born approximation, $\APV$ for a spin-zero nucleus is proportional to the ratio of weak ($\FW$) to charge ($\Fch$) form factors as~\cite{Horowitz:2001}:
\begin{equation}
    \APV = \frac{\sigma_R -\sigma_L}{\sigma_R+\sigma_L}\approx\frac{{\rm G}_{\rm F} \,Q^2}{4\pi\alpha\sqrt{2}}\frac{|Q_W|\, \FW(q)}{Z \, \Fch(q)},
\label{eqn:apv}
\end{equation}
where $\rm \sigma_R$ ($\rm \sigma_L$) is the elastic differential cross-section of right (left) handed electrons off the target with a four-momentum transfer squared Q$^2$, $q=\sqrt{Q^2}$. G$\rm _F$ is the Fermi constant, $\alpha$ is the fine structure constant, and the weak charge of $\eca$ is $Q_{\rm W} = -26.0\pm 0.1$ \cite{supplemental}. 
$\Fch$ from existing measurements \cite{Emrich:1983xb,supplemental} is used to extract $\FW$ from the measured $\APV$. 
The requirements for the practical application of this formula including precise Coulomb distortion calculations~\cite{Horowitz:1998vv} are described elsewhere~\cite{Horowitz:2001}. 

With the PREX-2 apparatus~\cite{PREX:2021umo} reoptimized to measure scattering from the calcium target, $\APV$ was measured at a four-momentum transfer just below the first diffractive cross-section minimum of $\eca$ to achieve high sensitivity to the neutron skin. 
Using two dipole magnets, 4$^{\circ}$--6$^{\circ}$ scattered electrons from a 2.18~GeV beam impinging on the calcium target were directed through precisely-machined collimators into the acceptance of the two High-Resolution Spectrometers (HRSs)~\cite{HallA_NIM} placed symmetrically 
on either side of the beam-axis. The elastically scattered electrons were focused into a peak with a momentum dispersion of about 16~m, and intercepted by a single Cherenkov detector in each HRS arm consisting of a  $16\,{\rm cm}\times3.5\,{\rm cm}\times0.5\,{\rm cm}$ fused-silica tile. Total internal reflection provided efficient Cherenkov light transmission to a photo-multiplier tube (PMT) coupled to the tile. The edge of the tile was positioned to ensure a momentum cut-off at $\sim$2~MeV below the elastic peak, thus minimizing contributions from inelastic scattering.

The polarized electron beam was generated using circularly polarized laser light incident on a photocathode~\cite{Source1}. The beam polarization sign follows the handedness of the laser circular polarization selected at 120~Hz using a Pockels cell, creating 8.13~ms time windows of constant beam helicity arranged in quartet patterns
($+--+$ or $-++-$) to ensure cancellation of 60~Hz AC power pickup. The sign of each quartet was selected pseudo-randomly and reported to the data acquisition system (DAQ) with a delay to suppress electronic pickup.

Production data totaling 412 Coulombs were acquired with a 150~$\mu$A beam rastered over a 4~mm$^2$ area on enriched $\eca$ targets mounted on a cryogenically  cooled copper ladder. Two $~1~{\rm g/cm^2}$ targets, with atomic $\eca$ percent of $95.99 \pm 0.02$\% and $91.70 \pm 0.01$\% were used to acquire 7.8\% and 92.2\% of the total data, respectively. 

The PMT anode current from the $\approx$28~MHz scattered flux in each detector was integrated and digitized over each helicity window by high-precision 18-bit sampling ADCs. 
The PMT was bench-tested before and after the run using light sources mimicking the integrated Cherenkov light response to determine linearity under operating conditions. Linearity was cross-checked throughout the run by monitoring detector output variation with beam current. The independent asymmetry measurements from each HRS were combined with equal weight; the final data set comprised 87M window quartets.

The beam intensity, energy and trajectory at the target  were measured with beam monitors using the same integrating DAQ. Three radio frequency (RF) cavities (BCMs) measured the beam intensity, while six RF antenna monitors (BPMs) measured beam position along the beam line, including at dispersive locations with energy sensitivity.
The polarized source was tuned to minimize the average helicity-correlated changes in beam parameters on target~\cite{Palatchi:2021vrh}.
Two techniques were used to reverse the beam polarization relative to the voltage applied to the Pockels cell.
A half-wave plate (HWP) was inserted in the laser beam path, separating the data sets into alternating reversal
states with a period of about ten hours.  The full production data set was additionally divided into three parts characterized by a change in spin precession in the low energy injector which reversed (or not) the polarization sign on target relative to that at the polarized source. Averaging over these reversals further suppressed spurious  helicity-correlated asymmetries in $\APV$. 

The helicity-correlated integrated beam charge asymmetry 
was controlled using active feedback, and averaged to -89~ppb over the run. Modulations of air-core magnets and an accelerating RF cavity placed upstream of all BPMs
were 
used to calibrate detector sensitivities.  This calibration was crosschecked with a regression analysis based on intrinsic beam fluctuations. 
The individual quartet measurements of $\APV$ were corrected for beam intensity, trajectory and energy fluctuations; the helicity-correlated correction averaged to $53\pm5$~ppb over the run. Consistency checks demonstrated that the residual detector asymmetry fluctuations were dominated by counting statistics.

Two  polarimeters measured the longitudinal beam polarization P$_{\rm b}$ upstream of the target. Operating continuously through the run, the Compton polarimeter used a calorimeter to measure the energy of photons scattered by the electron beam
traversing an optical cavity of circularly polarized green laser light~\cite{Friend:2011qh}. Calibration uncertainties were minimized by integrating the calorimeter response for each helicity window, thereby eliminating a low-energy threshold. Another polarimeter that detected M{\o}ller-scattered electrons from a polarized iron foil target in a 4~T magnetic field was deployed 9 times periodically during the run. The results were consistent between polarimeters and combined to yield P$\rm _b =87.10\ \pm\ $0.39\%.

Calibration data were collected at reduced beam current (100~nA to 1$~\mu$A) to enable counting and tracking of individual electrons. With Cherenkov detector PMT gains increased to detect individual particle pulses in coincidence with drift chamber tracks and trigger scintillators hits, the reconstructed scattering angle and momentum were calibrated using scattered electrons from a thin carbon target and a steel-walled water flow target, mounted on a separate, water-cooled target ladder. The momentum recoil difference between elastic scattering from hydrogen and oxygen in the water target calibrates the central angle to 0.02$^\circ$ absolute accuracy.  

Similar counting data collected with the production $\eca$ target were used to estimate the fractional contribution from the first three low-lying excited states in $\eca$, which totaled 1.4\% of the accepted rate. Calculation of the excited state asymmetries and conservative uncertainties~\cite{supplemental} lead to the $\APV$ corrections listed in Table~\ref{tab:systematics}. 
The $\eca$ parity-conserving transverse single-spin asymmetry $\AT$ was independently measured~\cite{PREX:2021uwt} and, along with counting data, used to estimate a 13~ppb uncertainty in the $\AT$ correction to $\APV$, due to potential residual transverse beam polarization coupled to imperfect symmetry in the left-right and top-bottom acceptance.

\begin{figure}[ht]
    \centering
    \includegraphics[width=0.45\textwidth]{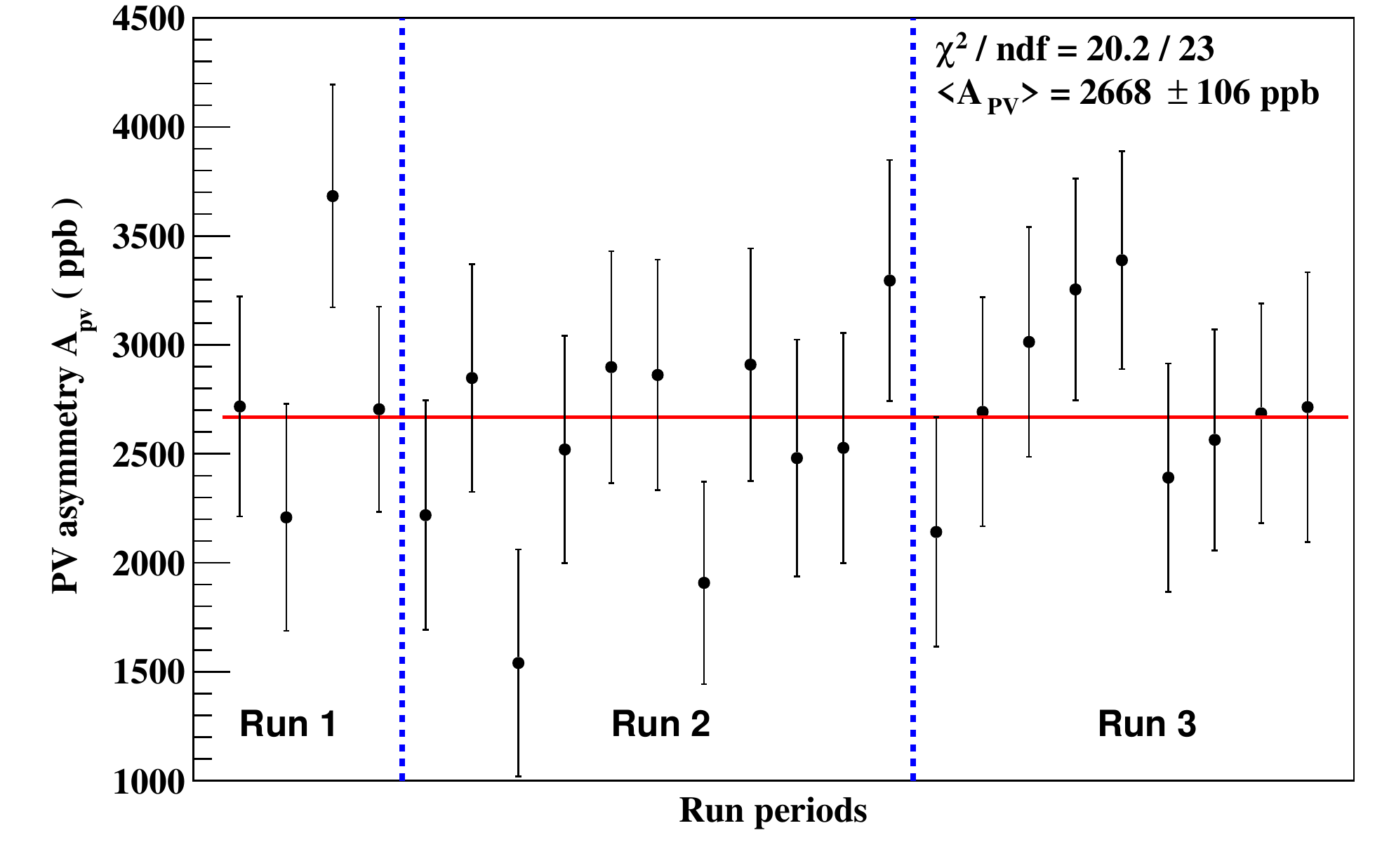}
    \caption{Measurements of $\APV$ with statistical uncertainty; each $\approx$40~hour period includes two states with complementary HWP settings. The three run periods demarcate injector spin orientation reversals. }
    \label{fig:Pittplot}
\end{figure}

Using a theoretically computed $\APV(^{40}$Ca$) = 2430 \pm 30$~ppb~\cite{supplemental}, the $\APV$ contribution from the assayed 7.95\% $^{40}$Ca target fraction was calculated to be $19 \pm 3$~ppb.
Figure~\ref{fig:Pittplot} shows $\APV$ measurements after all corrections in roughly uniform periods, with the global average $\APV = 2668\pm106 ~{\rm ppb}$. 

To compare this result to a theoretical model, 
the acceptance function $\epsilon(\theta)$ provides the distribution of scattering angles intercepting the Cherenkov detectors:
\begin{equation}
     \langle {\rm A} \rangle = \frac{\int d\theta\, \sin\theta\,{\rm A}(\theta) \, \frac{d\sigma}{d\Omega}\,\epsilon(\theta)}{\int d\theta\,\sin\theta\,\frac{d\sigma}{d\Omega}\,\epsilon(\theta)}
 \end{equation}
where $\frac{d\sigma}{d\Omega}$ is the differential cross-section and A($\theta$) is the modeled parity violating asymmetry as a function of scattering angle~\cite{datafile}. 
Simulation modeling of the calibration data was used to calculate $\epsilon(\theta)$.  Radiative and rescattering effects in the target
 change the average accepted angle by 1.5\%. 
The mean kinematics were found to be  $\langle \theta \rangle$ = 4.51$^{\circ} \pm 0.02^{\circ}$ and $\langle Q^2\rangle= 0.0297~\pm~0.0002~ {\rm (GeV/c)}^2$. Alternative acceptance functions, calculated using geometric and magnetic tolerances but still constrained to match spectra from calibration runs, were used to calculate an uncertainty of $\pm 24$~ppb on $\APV$ due to possible variation of $\epsilon(\theta)$.

\begin{table}[ht]
 \caption{$\APV$ corrections and corresponding systematic uncertainties, normalized to account for polarization and background fractions.}
    \label{tab:systematics}
    \begin{center}
\begin{tabular}{l D{,}{\ \pm\ }{-1} D{,}{\ \pm\ }{-1}}
\toprule
Correction                & \multicolumn{1}{c}{Absolute [ppb]}   & \multicolumn{1}{c}{Relative [\%]}  \\
\midrule
Beam polarization & 382 ,  13 & 14.3 , 0.5 \\
Beam trajectory \& energy & 68 , 7 & 2.5 , 0.3 \\
Beam charge asymmetry & 112 , 1 & 4.2 , 0.0 \\
Isotopic purity & 19 , 3 & 0.7 , 0.1 \\
3.831 MeV (2$^+$) inelastic & -35 , 19 & -1.3 , 0.7 \\
4.507 MeV (3$^-$) inelastic & 0 , 10 & 0 , 0.4 \\
5.370 MeV (3$^-$) inelastic & -2 , 4 & -0.1 , 0.1 \\
Transverse asymmetry & 0 , 13 & 0 , 0.5 \\
Detector non-linearity & 0 , 7 & 0 , 0.3 \\
Acceptance & 0 , 24 & 0 , 0.9 \\
Radiative corrections ($Q_W$) & 0 , 10 & 0 , 0.4 \\ 
\midrule
Total systematic uncertainty & \multicolumn{1}{r}{40 ppb} & \multicolumn{1}{r}{1.5\%} \\
Statistical uncertainty & \multicolumn{1}{r}{106 ppb} & \multicolumn{1}{r}{4.0\%} \\
\bottomrule
\end{tabular}
\end{center}
\end{table}

Table~\ref{tab:systematics} lists all significant corrections and corresponding uncertainties; the total systematic uncertainty is 40~ppb. 

The weak form factor is directly related to $\APV$ in Eqn~\ref{eqn:apv}, and is the Fourier transform of the weak charge density $\rho_{\rm W}$,  
\begin{equation}
    \FW(q)=\frac{1}{Q_{\rm W}}\int d^3r\, j_0(qr) \rho_{\rm W}(r).
    \label{eq.FW}
\end{equation}
We assume a shape for $\rho_W(r)$ and calculate $\APV$,  including Coulomb distortions and integrating over the acceptance $\epsilon(\theta)$.  After adjusting the radius parameter in the  $\rho_W(r)$ model~\cite{supplemental} to reproduce the measured $\APV$, we evaluate $\FW(q)$ in Eq. \ref{eq.FW} using this $\rho_W(r)$ at the reference momentum transfer $q=0.8733$ fm$^{-1}$.  This procedure is insensitive to the form of the model $\rho_W$ and yields the results in Table \ref{Tab.results}.


\begin{table}[ht]
 \caption{CREX form factor results for $\eca$, with $q$ and $\Fch$ input values. The uncertainties are due to statistics and experimental systematics, respectively.}
    \label{Tab.results}
\begin{tabular}{c l c l c l} 
    \toprule
    Quantity  & Value & $\pm$ & (stat) & $\pm$ & (sys) \\
    \midrule
    $q$ &  \multicolumn{5}{l}{0.8733 fm$^{-1}$}\\
    $\FW(q)/\Fch(q)$~~~ & 0.8248 & $\pm$ & 0.0328 & $\pm$  & 0.0124\\
    $\Fch(q)$ &   \multicolumn{5}{l}{0.1581}\\
    $\FW(q)$ & 0.1304 & $\pm$ & 0.0052 & $\pm$ & 0.0020 \\
    $\Fch(q)-\FW(q)$ & 0.0277 &$\pm$& 0.0052& $\pm$& 0.0020\\
     \bottomrule    
    \end{tabular}
\end{table}

While the extracted value of $\FW$ depends on $\Fch$,
$\FW/\Fch$ and $\Fch-\FW$ are quite insensitive to  $\Fch$. 
In order to determine $\Fch (q)=\int d^3r\, j_0(qr)\rho_{\rm ch}(r)/Z$, we use a composite charge density for $\eca$ starting with an accurate sum of Gaussians density for $^{40}$Ca \cite{SICK1979245} and add a Fourier Bessel expansion for the small difference between the charge densities of $\eca$ and $^{40}$Ca \cite{Emrich:1983xb,DeVries:1987}, see~\cite{supplemental}.  This procedure yields a $\eca$ charge radius of 3.481 fm, close to the experimental value of 3.477 fm  \cite{LandoltBornstein2004:sm_lbs_978-3-540-45555-4_22}. 

A main result of this paper is a measurement of the 
difference between charge and weak form factors,
\begin{equation}
    \Fch(q)-\FW(q)=0.0277 \pm 0.0055\  ({\rm exp}).
    \label{eq.FchW}
\end{equation}
 The uncertainty is the quadrature sum of the experimental statistical and systematic uncertainties, referred to henceforth as the experimental error (exp), dominated by counting statistics. We emphasize that the Eq.~\ref{eq.FchW} result is model-independent and quite insensitive to the assumed shape for the weak density $\rho_W(r)$.

 \begin{figure}[htb]
    \centering
    \includegraphics[width=0.48\textwidth]{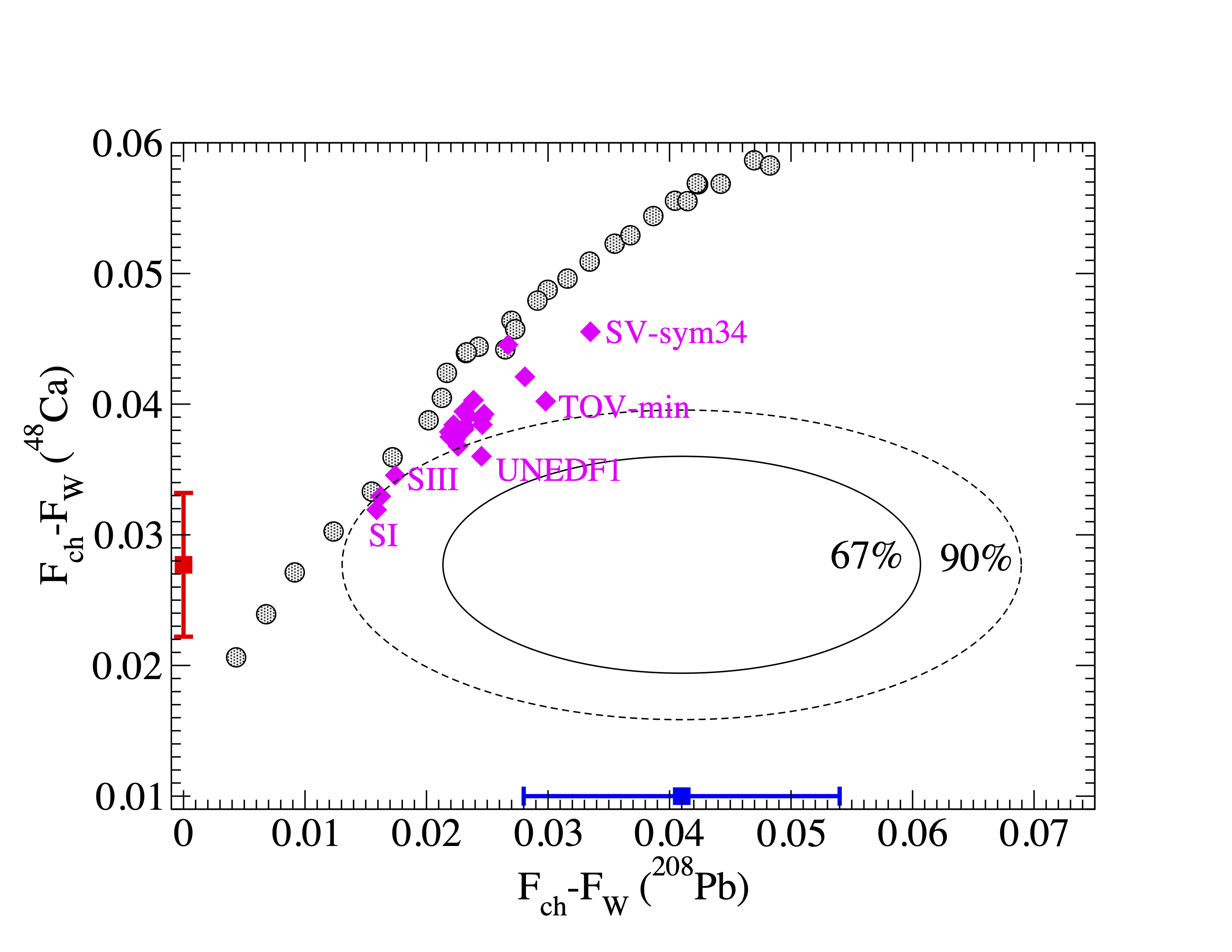}
    \caption{Difference between the charge and weak
    form factors of $^{48}$Ca (CREX) versus that of $^{208}$Pb (PREX-2) at their respective momentum transfers.  The blue (red) data point shows the PREX-2 (CREX) measurements. The ellipses are joint PREX-2 and CREX 67\% and 90\% probability contours.  The gray circles (magenta diamonds) are a range of relativistic (non-relativistic) density functionals.  For clarity only some of these functionals are labeled.  The complete list is in ref. \cite{supplemental}. }
    \label{PREXVsCREX}
\end{figure}
 
Figure \ref{PREXVsCREX} displays Eq.~\ref{eq.FchW} for $^{48}$Ca along with the \mbox{PREX-2} result $\Fch-\FW=0.041\pm0.013$ for $^{208}$Pb at a smaller momentum transfer of 0.3977~fm$^{-1}$ \cite{PREX:2021umo}.  The figure also shows a series of relativistic energy functional models with density-dependent symmetry energy slope parameter $L$~\cite{HOLT201877,Lsym} that varies over a large range from small negative values at the lower left to large positive values at the upper right. Additionally, a diverse collection of non-relativistic density functional models are shown~\cite{supplemental}. 
Here, $\Fch$ and $\FW$ include proton and neutron densities folded with single nucleon electric and magnetic form factors and spin orbit currents \cite{PhysRevC.86.045503}.
The models that best reproduce both the CREX and PREX-2 results tend to predict $\Fch-\FW$ slightly below the PREX-2 result for $^{208}$Pb and slightly above the CREX result for $^{48}$Ca. 
 
\begin{figure}[htb]
    \centering
    \includegraphics[width=0.48\textwidth]{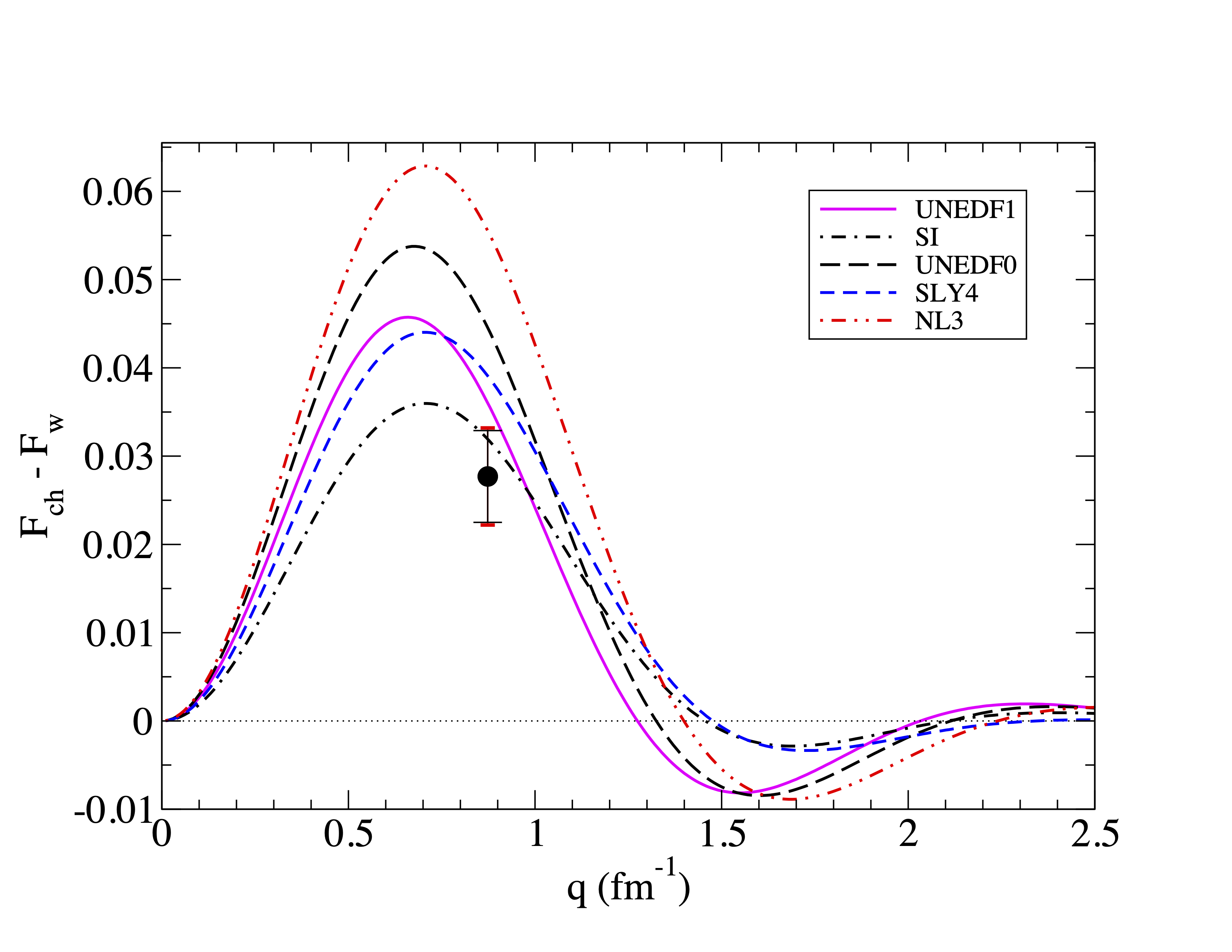}
    \caption{The difference between the charge and weak form factors for $^{48}$Ca as a function of momentum transfer q = $\sqrt{Q^2}$. The curves show results for non-relativistic (SI, SLY4, UNEDF0, UNEDF1) and relativistic (NL3) density functional models. The CREX measurement is indicated by a circle with the inner black error bar showing the contribution from statistics and the total experimental error bar in red.}
    \label{QweakVsQ}
\end{figure}

Figure \ref{QweakVsQ} shows the momentum transfer dependence of $\Fch-\FW$ as predicted by a few non-relativistic and relativistic density functional models.  
It is evident that some model results cross as a function of $q$, emphasizing the somewhat different $q$ dependence. In the limit $q\rightarrow 0$, $\Fch(q)-\FW(q) \approx q^2(\RW^2-\Rch^2)/6$, where $\RW$ is the rms radius of $\rho_W(r)$ and $\Rch$ is the charge radius.  Since this equation is not valid at the larger $q$ of CREX, the extraction of $\RW-\Rch$ introduces some model dependence.  

\begin{figure}[htb]
    \centering
    \includegraphics[width=0.48\textwidth]{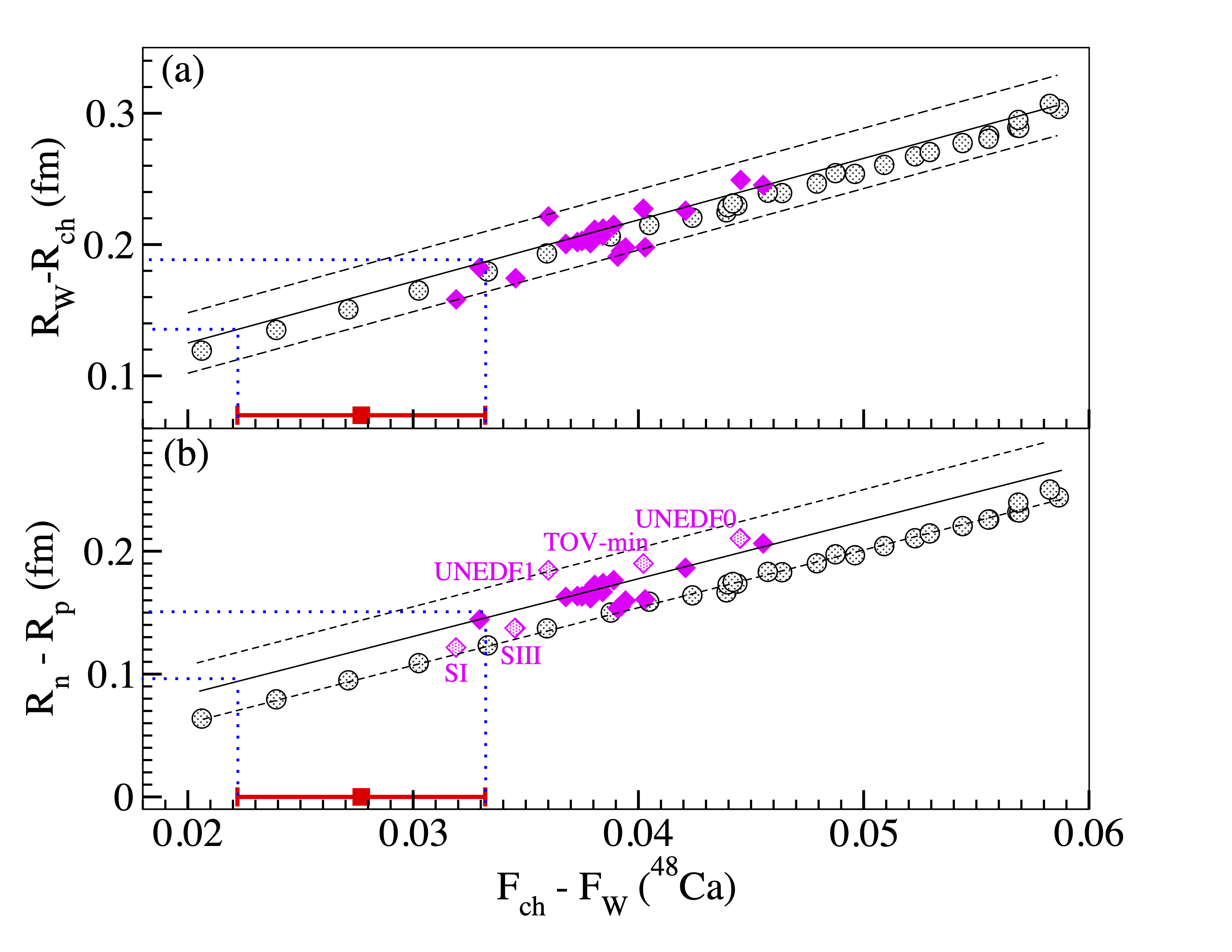}
    \caption{(a) $^{48}$Ca weak minus charge rms radius versus charge minus weak form factor at the CREX momentum transfer.  The CREX experimental value and uncertainty is shown (red square).  The gray circles (magenta diamonds) show a range of relativistic (non-relativistic) density functionals. (b) $^{48}$Ca neutron minus proton rms radius  versus charge minus weak form factor. 
    } 
    \label{Rnp_FF:fig}
\end{figure}

\begin{figure}[htb]
    \centering
    \includegraphics[width=0.48\textwidth]{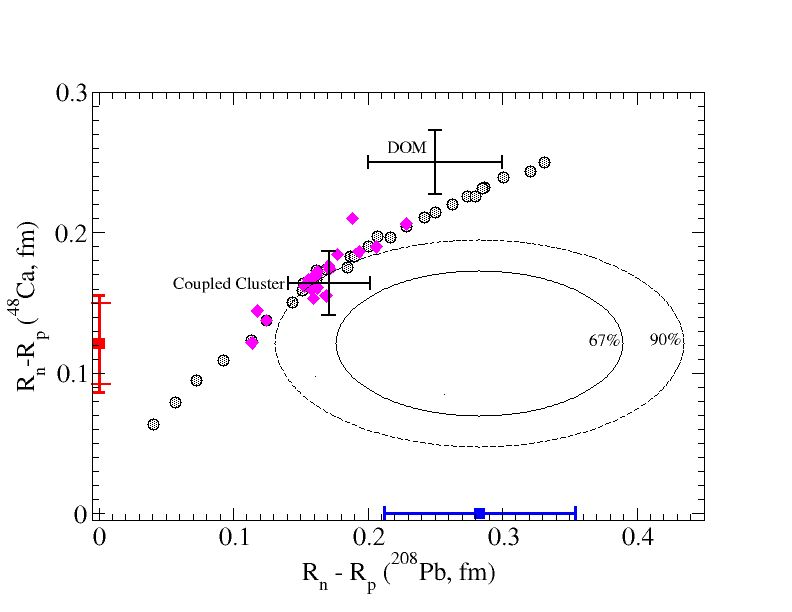}
    \caption{$^{48}$Ca neutron minus proton radius versus that for $^{208}$Pb.  The PREX-2+PREX-1 experimental result is shown as a blue square, while that for CREX is shown as a red square with the inner error bars indicating the experimental error and the outer error bars including the model error.  
    The gray circles (magenta diamonds) show a variety of relativistic (non-relativistic) density functionals.  Coupled cluster \cite{Hu:2021} and dispersive optical model (DOM) predictions \cite{PhysRevC.101.044303} are also shown.
}
    \label{RnpCREX_PREX:fig}
\end{figure}

Relativistic and non-relativistic density functional model predictions of $\RW-\Rch$ are plotted in 
Fig.~\ref{Rnp_FF:fig}(a) versus $\Fch(q)-\FW(q)$ . The somewhat different $\rho_W(r)$ shapes lead to the vertical spread in the non-relativistic models. 
Figure \ref{Rnp_FF:fig}(b) shows a similar plot of point neutron minus proton radii $R_n-R_p$ versus $\Fch(q)-\FW(q)$.  
To calculate $R_n-R_p$ given $\Fch-\FW$ one must include full current operators including spin orbit ($\vec{L}\cdot\vec{S}$) contributions [41].  Relativistic models tend to have somewhat larger $\vec{L}\cdot\vec{S}$ currents.  As a result the gray circles in Fig.~\ref{Rnp_FF:fig}(b) are somewhat lower than those in Fig.~\ref{Rnp_FF:fig}(a) when compared to non-relativistic models.
Lines with slope matching that of the relativistic model variation are drawn to enclose the full range of displayed models, providing the model range and central values 
listed in Table \ref{Tab.RwchRnp}.
This underscores the fact that the CREX $^{48}$Ca $R_n-R_p$ has significant modeling uncertainty, in contrast to the PREX $^{208}$Pb $R_n-R_p$, see \cite{supplemental}.  Reduced model uncertainty would result if theoretical predictions were compared to the model-independent $\Fch-\FW$ in Fig. \ref{PREXVsCREX} rather than to $R_n-R_p$ in Fig. \ref{RnpCREX_PREX:fig}.  

\begin{table}[htb]
 \caption{Extracted $\RW-\Rch$ and $R_n-R_p$ radii.  The first uncertainty is experimental and the second reflects the shape uncertainty in $\rho_W(r)$ estimated from the spread in Fig. \ref{Rnp_FF:fig}.}
    \label{Tab.RwchRnp}.

    \begin{tabular}{c c}
    \toprule
    Quantity ~~~& Value $\pm$ (exp) $\pm$ (model) [fm] \\
    \midrule
   $\RW-\Rch$ & $0.159 \pm 0.026 \pm 0.023$\\
   $R_n-R_p$ & $0.121 \pm 0.026 \pm 0.024$     \\
\bottomrule
    \end{tabular}

\end{table}

 $R_n-R_p$ for $^{48}$Ca versus $R_n-R_p$ for $^{208}$Pb is shown in Fig. \ref{RnpCREX_PREX:fig}.  
 A number of models including the microscopic coupled cluster calculations~\cite {Hu:2021} are consistent with our results, slightly under-predicting $^{208}$Pb while slightly over-predicting $^{48}$Ca. Dispersive optical model    
calculations agree well for $^{208}$Pb but substantially over-predict $R_n-R_p$ for $^{48}$Ca \cite{PhysRevC.101.044303}.

In conclusion, we have reported a new and precise measurement of the PVES asymmetry from $^{48}$Ca and a model-independent extraction of the difference between the charge form factor and the weak form factor $\Fch - \FW$ at $q = 0.8733 \hskip 0.03in {\rm fm}^{-1}$.  
In addition, we have extracted, with a small model dependence, the weak skin $\RW - \Rch$ and the neutron skin $R_n - R_p$ of $^{48}$Ca 
and compared it to that of $^{208}$Pb. 
The extracted neutron skin of $^{48}$Ca (CREX) is relatively thin compared to the  prediction of most models, while that of $^{208}$Pb (PREX) 
is thick, yet both are consistent with a number of density functional models and with the microscopic coupled cluster models~\cite {Hu:2021}. This will have implications for future energy density functional calculations and the density dependence of the symmetry energy.

The small model dependence of this result could be further constrained with a future measurement of $\APV$ from $\eca$ at an additional $Q^2$~\cite{PhysRevC.102.064308}. Experimental techniques from this work, including excellent systematic control of helicity-correlated fluctuations and demonstration of high precision electron beam polarimetry, will inform the design of future projects MOLLER \cite{MOLLER_experiment} and SoLID \cite{SOLID_proposal} at JLab measuring fundamental electroweak couplings, as
well as P2 and the $^{208}$Pb radius experimental proposals at Mainz \cite{Mainz_P2}. 

We thank the entire staff of JLab for their efforts to
develop and maintain the polarized beam and the experimental 
apparatus, and acknowledge the support of the U.S.
Department of Energy, the National Science Foundation
and NSERC (Canada). We thank J. Piekarewicz, P. G. Reinhard and X. Roca-Maza for RPA calculations of $^{48}$Ca excited states and J. Erler and M. Gorchtein for calculations of $\gamma-Z$ radiative corrections. This material is based upon the
work supported by the U.S. Department of Energy, Office
of Science, Office of Nuclear Physics Contract No. DE-
AC05-06OR23177.

\bibliography{CREX,references,prex2}

\end{document}